\documentclass[aps,prd]{revtex4}
\usepackage{epsfig}
\newcommand{\vlamb}{\mbox{\boldmath$\lambda$\unboldmath}}
\newcommand{\vrho}{\mbox{\boldmath$\rho$\unboldmath}}

\newcommand{\be}{\begin{equation}}
\newcommand{\ee}{\end{equation}}
\newcommand{\bea}{\begin{eqnarray}}
\newcommand{\eea}{\end{eqnarray}}
\newcommand{\bean}{\begin{eqnarray*}}
\newcommand{\eean}{\end{eqnarray*}}

\newcommand{\gapproxeq}{\lower
.7ex\hbox{$\;\stackrel{\textstyle >}{\sim}\;$}}
\newcommand{\lapproxeq}{\lower
.7ex\hbox{$\;\stackrel{\textstyle <}{\sim}\;$}}

\newcommand{\nrightarrow}{\mbox{$ \rightarrow\hspace{-0.4cm}\backslash\hspace{0.4cm}$}}
\def\3bar{$\bar {\hbox{\bf 3}}$}

\begin{document}
\bibliographystyle{unsrt}

\title{\bf Quarks, diquarks and QCD mixing in the $N^*$ resonance spectrum~\footnote{Dedicated to 
the memory of R.H. Dalitz.}}

\author{Qiang Zhao$^{1,2}$\footnote{e-mail: Qiang.Zhao@surrey.ac.uk}
 and  Frank E. Close$^3$\footnote{e-mail: F.Close@physics.ox.ac.uk}}
\affiliation{1) Institute of High Energy Physics, Chinese Academy
of Sciences, Beijing, 100049, P.R. China}
\affiliation{2)
Department of Physics, University of Surrey, Guildford, GU2 7XH,
United Kingdom}
\affiliation{3) Rudolf Peierls Centre for
Theoretical Physics, University of Oxford, Keble Rd., Oxford, OX1
3NP, United Kingdom}

\date{\today}

\begin{abstract}
We identify a ``$\Lambda$ selection rule" for $N^*$ resonances in
the presence of QCD mixing effects. We quantify these mixing
effects from existing data and predict amplitudes for exciting
{\bf 20} representations in SU(6), which are forbidden in strict
diquark models. By classifying Particle-Data-Group (PDG) states at
$N=2$, we show that $\gamma N\to K\Lambda$, $K^*\Lambda$,
$K\Sigma$, $K^*\Sigma$, and $J/\psi \to \bar{p}N^*$ are ideal
probes of baryon dynamics and for establishing whether strongly
correlated diquarks survive for $L > 0$.

\end{abstract}

\maketitle


\vskip 1.cm

\section{Introduction}

It is remarkable that forty years after the quark model was first
applied to the problem of baryon resonances~\cite{dalitz} it is
still not well established whether three constituent quarks are
the minimal effective degrees of freedom or whether a
quark-diquark dynamics, where a pair of quarks is ``frozen" into a
ground state, suffices. Indeed, the vast literature inspired by
the apparent discovery of a metastable ``pentaquark" baryon in
2003~\cite{penta} showed both how little the strong dynamics of
quarks is understood and raised renewed speculation about the role
and existence of highly correlated diquarks~\cite{jw,kl}.
Furthermore, an acceptable description of baryon resonance
spectroscopy has been proposed based on a quark-diquark
picture~\cite{wilc04}.

The relative coordinate between the two quarks forming the diquark
is constrained to be in the $l_{\rho}=0$ state. (We use the
symbols $\rho,\lambda$ as in Ref~\cite{capstick-roberts} to denote
the antisymmetric and symmetric two-body substates within a
three-body wavefunction). In such a case, the familiar and
established $[SU(6), L^P]$ multiplets [{\bf 56},$L^P$] and [{\bf
70},$L^P$] occur, but it is impossible to form $[SU(6), L^P]$
correlations, [{\bf 20},$L^P$]~\cite{licht}. Within a $qqq$
dynamics, where both $\lambda$ and $\rho$ spatial oscillators can
be excited, the spectrum is richer and such {\bf 20} states can
occur.

Whereas [{\bf 56},$L^P$] and [{\bf 70},$L^P$] excitations are well
established, the search for [{\bf 20},$L^P$] has been largely
ignored, primarily because they cannot be excited by mesons or
photons from a {\bf 56} nucleon. This is because photons and
$q\bar{q}$ beams transform under SU(6) as {\bf 35}, which with the
SU(6) forbidden transition {\bf 56}$\otimes${\bf 35}
$\nrightarrow$ {\bf 20} causes them to decouple from nucleons in
naive SU(6) ~\cite{licht,faiman}.

The purpose of this paper is to re-examine the assumptions
underlying resonance production in the quark model. This will lead
us to a selection rule, that appears to have been overlooked in
the literature, and also to identify circumstances where {\bf 20}
states can be excited.

A standard and phenomenologically successful assumption common to
a large number of papers in the quark model is that photon
transitions are additive in the constituent
quarks~\cite{close-book,cko,fkr}. This assumption also underlies
models of hadronic production and decay in the sense that when
$q_1q_2q_3 \to  [q_1q_2q_i] + [q_3 \bar{q_i}]$, the quark pair
$q_1q_2$ are effectively spectators and only $q_3$ is involved in
driving the transition. Such approximations lead to well known
selection rules, which have proved useful in classifying
resonances\cite{close-book}. We adopt this approximation as a
first step and show that within it there is a further selection
rule that appears to have been overlooked in the literature. We
shall refer to this as the ``$\Lambda$ selection rule" and show
how it may help classify $N^*$ resonances.

The above ``spectator-hypothesis" for transition amplitudes will
be violated by the spin-dependent forces that act between pairs of
quarks and break SU(6), such as those generating the $N$-$\Delta$
mass gap. For example, when the nucleon is in an electromagnetic
field, gauge invariance and the presence of such two-body forces
imply that there occur diagrams where the photon interacts with
quark number 3, say, and the exchange force acts between quarks 3
and 1 or 2. Thus electromagnetic interactions can transfer
momentum to both $\lambda$ and $\rho$ oscillators, which both
spoils the spectator-hypothesis and opens the possibility of
exciting {\bf 20}-plets.

As a specific and quantifiable example we shall assume these
spin-dependent forces arise from gluon-exchange in QCD. This has
considerable quantitative support~\cite{dgg} and also has been
shown to induce mixings, including {\bf 70}-plet configurations,
into the nucleon wavefunction~\cite{ikk}. Taking into account that
the $N$-$\Delta$ mass gap of $300$ MeV is on the scale of
$\Lambda_{QCD}$, the resulting mixing effects can be sizeable.
Whereas {\bf 56}$\otimes${\bf 35} $\nrightarrow$ {\bf 20},
 the coupling {\bf 70}$\otimes${\bf 35} $\rightarrow$ {\bf 20} is allowed.
Consequently the SU(6) breaking that induces {\bf 70} correlations
into the nucleon enables the excitation of {\bf 20}-plets by
photons and mesons. Interestingly, we shall see that the $\Lambda$
selection rule still manifests itself in $N^*\to K\Lambda$
transitions even though the SU(6) symmetry is broken. This
phenomenon will be useful for clarifying the excitation of {\bf
20}-plets in the $N^*$ spectrum.

In the next section we first present the ``$\Lambda$ selection
rule". We then
 show how QCD-generated wavefunction mixing allows production of {\bf 20}-plets.
 We formulate these ideas in a
QCD quark model~\cite{ikk} though the qualitative results should
be more generally true. In the final section we quantify these
effects and discuss their application to $N^*$ classification.

\section{The Model}

For reference, we specify our nomenclature. The standard
SU(6)$\otimes$O(3) wavefunction can be constructed from three
fundamental representations of group $S_3$:
\begin{eqnarray}
SU(6): \ \ & {\bf 6}\otimes{\bf 6}\otimes{\bf 6} &= {\bf 56}_s
+{\bf 70}_\rho + {\bf 70}_\lambda + {\bf 20}_a ,
\end{eqnarray}
where the subscripts denote the corresponding $S_3$ basis for each
representation, and the bold numbers denote the dimension of the
corresponding representation. The spin-flavor wavefunctions
can be expressed as $|{\bf N}_6, ^{2S+1} {\bf N}_3\rangle $, where
${\bf N}_6$ (={\bf 56, \ 70} or {\bf 20}) and ${\bf N}_3$ (={\bf
8, \ 10}, or {\bf 1}) denote the SU(6) and SU(3) representation
and $S$ stands for the total spin. The SU(6)$\otimes$O(3) (symmetric) wavefunction is
\begin{equation}
|\mbox{SU(6)}\otimes\mbox{O(3)}\rangle = |{\bf N}_6, \
^{2S+1}{\bf N}_3, \ N, L, J\rangle \ ,
\end{equation}
where explicit expressions follow the convention of Isgur and
Karl~\cite{isgur-karl,ik-78,ik70mix}.

The basic rules follow from application of the Pauli exclusion
principle to baryon wavefunctions together with an empirically
well tested assumption that electroweak and strong decays are
dominated by single quark transitions where the remaining two
quarks, or diquark, are passive
spectators~\cite{capstick-roberts}. In particular, selection rules
for specific processes can resolve the underlying dynamics. For
example, the Moorhouse selection rule~\cite{moorhouse} states that
transition amplitudes for $\gamma p$ to all resonances of
representation $[{\bf 70}, \ ^4 8]$, such as $D_{15}(1675)$, must
be zero due to the vanishing transition matrix element for the
charge operator.

Such correlations also lead to a  ``$\Lambda$ selection rule",
which appears to have been overlooked in the literature. It states
that $N^*$ in [{\bf 70}, \ $^4 8$] decouple from $\Lambda K$ and
$\Lambda K^*$ channels. This follows because the $[ud]$ in the $\Lambda$ has
$S=0$ and in the spectator approximation,  the
strangeness emissions in $N^* \to \Lambda K$ or $\Lambda K^*$,
 the spectator $[ud]$ in the $N^*$ must also be in
$S_{[ud]}=0$, whereby such transitions for the $N^*$ of $[{\bf 70},
\ ^4 8]$ with $S_{[ud]}=1$ are forbidden.

Note that the $\Lambda$ selection rule applies to both proton and
neutron resonances of $[{\bf 70}, ^4 8]$, in contrast to the
Moorhouse selection rule, which applies only to the proton. The
nearest that we can find to this in the literature is that
$\Lambda^*[{\bf 70}, \ ^4 8] \nrightarrow \bar{K}N$~\cite{hey}.
While the associated zero in $N^*[{\bf 70}, \ ^4 8] \rightarrow K
\Lambda$ is implicit in work that has calculated the couplings of
baryon resonances~\cite{capsr1}, the source and generality of the
rule does not seem to have been noted~\cite{fai77}.

\section{Application to $N^*$ spectrum}

An immediate application of the rules is to the $D_{15}(1675)$,
which is in $[{\bf 70}, ^4 8]$. According to the Moorhouse
selection rule, the amplitudes for $\gamma p \to D_{15}$ should
vanish. However, the experimental values are not zero, though they
are small.  Non-zero amplitudes arise from QCD mixings induced by
single gluon exchange in the physical nucleon~\cite{ikk}. The
effective interaction
\begin{equation}
H_{FB} = \frac{2\alpha_s}{3m_i m_j} \left[\frac{8\pi}{3}{\bf
S}_i\cdot{\bf S}_j\delta^3({\bf r}_{ij})
+\frac{1}{r_{ij}^3}\left(\frac{3({\bf S}_i\cdot {\bf r}_{ij})({\bf
S}_j\cdot {\bf r}_{ij})}{r_{ij}^2}  - {\bf S}_i\cdot{\bf S}_j
\right) \right] \label{hfb}
\end{equation}
induces significant mixings between the $^2${\bf 8} and $^4 ${\bf
8} in the {\bf 56} and {\bf 70}~\cite{ik70mix} and the nucleon
wavefunction becomes~\cite{ikk}
\begin{equation}
|N\rangle = 0.90|^2S_S\rangle -0.34|^2S_{S'}\rangle -0.27|^2S_M
\rangle -0.06|^4 D_M\rangle \ , \label{nmix}
\end{equation}
where subscripts, $S$ and $M$, refer to the spatial symmetry in
the $S$ and $D$-wave states for the nucleon internal wavefunction.
Thus, the $O(\alpha_s)$ admixtures at $N=2$ comprise a 34\% in
amplitude excited {\bf 56} and 27\% {\bf 70} each with $L=0$ and
6\% {\bf 70} with $L=2$.  The {\bf 70} admixture quantitatively
agrees with the most recent data~\cite{pdg2004} for the $\gamma p
\to D_{15}$ amplitudes, neutron charge radius and $D_{05} \to
\bar{K} N$~\cite{ikk}. The results assume that mixing effects in
the $D_{15}$ are negligible relative to those for the
nucleon\cite{ikk}: this is because there is no $[{\bf 70}, ^2 8; L^P = 1^-]$ state
available for mixing with the $D_{15}$, and the nearest $J=5/2$
state with negative parity is over 500 MeV more massive at $N=3$.
Within this $O(\alpha_s)$ analysis, such mixing is negligible:
transitions from the large components of the nucleon to small in
$D_{15}$ have $\Delta N = 3$. The leading $O(\alpha_s)$ amplitude
for $\gamma p \to D_{15}$ is dominantly driven by the small
components in the nucleon and the large component in the
$D_{15}$~\cite{ik-78} for which $\Delta N =1$.

This violation of the Moorhouse selection rule supports the
hypothesis of QCD mixing in the wavefunction of the $N$. The
$\Lambda$ selection rule also remains robust, in the context of
the diquark model, as admixtures of $[ud]$ with spin one, which
would violate it, are only expected at most to be 20\% in
amplitude~\cite{ik-80}, to be compared with 27\% for the nucleon
in Eq.~(\ref{nmix}). Therefore, we expect the $\Lambda$ selection
rule to be at least as good as the Moorhouse rule even at
$O(\alpha_s)$. Thus decays such as $D_{15} \to K\Lambda$ will
effectively still vanish relative to $K\Sigma$; for the
$D_{15}(1675)$ the phase space inhibits a clean test but the ratio
of branching ratios for the analogous state at $N=2$, namely
$F_{17}(1990) \to K\Lambda : K \Sigma$, may provide a measure of
its validity. Secondly, for $\gamma n\to D_{15}$, where the
Moorhouse selection rule does not apply, the amplitudes are
significantly large and consistent with experiment~\cite{pdg2004}.
However, due to the $\Lambda$ selection rule, the $D_{15}^0
\nrightarrow K^0\Lambda$ which makes the search for the $D_{15}$
signals in $\gamma N\to K\Lambda$ interesting. An upper limit of
$B.R.< 1\%$ is set by the PDG~\cite{pdg2004} which in part may be
due to the limited phase space; a measure of the ratio of
branching ratios for $K \Lambda : K \Sigma$ would be useful. The
$F_{17}(1990)$, which is the only $F_{17}$ with $N=2$, is an ideal
candidate for such a test, which may be used in disentangling the assignments
of the positive parity $N^*$ at the $N=2$ level.

The QCD admixture of $[{\bf 70}, ^2 8, 2, 0, 1/2]$ in the nucleon
wavefunction enables the excitation of {\bf 20}-plets. There has
been considerable discussion as to whether the attractive forces
of QCD can cluster $[ud]$ in color {\bf $\bar{3}$} so tightly as
to make an effective bosonic ``diquark" with mass comparable to
that of an isolated quark. Comparison of masses of $N^*(u[ud])$
and mesons $u\bar{d}$ with $L \geq 1$ support this hypothesis of a
tight correlation, at least for excited
states~\cite{wilc04,fecscot}. If the quark-diquark dynamics is
absolute, then SU(6)$\otimes$O(3) multiplets such as $[{\bf
20},1^+]$ cannot occur. The spatial wavefunction for {\bf 20}
involves both $\rho$ and $\lambda$ degrees of freedom; but for an
unexcited diquark, the $\rho$ oscillator is frozen. Therefore,
experimental evidence for the excitations of the {\bf 20} plets
can distinguish between these prescriptions.

Pauli symmetry requires an antisymmetric spatial wavefunction for
a {\bf 20} e.g. for the lowest state $[{\bf 20}, 1^+]$
\be
\psi^a_{211}(\vrho,\vlamb)
=\sqrt{2}(\rho_+\lambda_z-\lambda_+\rho_z)\frac{\alpha_h^5}{\pi^{3/2}}
e^{-\alpha_h^2(\vrho^2+\vlamb^2)/2},
\ee
where $\rho_\pm\equiv \mp (\rho_x\pm i\rho_y)/\sqrt{2}$ and
$\lambda_\pm\equiv \mp (\lambda_x\pm i\lambda_y)/\sqrt{2}$, and
hence both $\rho$ and $\lambda$ degrees of freedom have to be
excited. Furthermore {\bf 20} decouples from {\bf 35}$\otimes${\bf
56} but is allowed to {\bf 35}$\otimes${\bf 70}. A 27\% {\bf 70}
admixture in the nucleon has potential implications for resonance
excitation that may be used to look for {\bf 20}-plets.

For transitions between representation $[{\bf 70}, ^2 8, 2, 0, 1/2
]$ and $[{\bf 20}, ^2 8, 2, 1, J]$ the matrix element
$\langle\psi_{210}^a | e^{ik r_{3z}}|\psi_{200}^\rho\rangle \equiv
0$. Non-zero photon transitions can occur by the orbital flip
``electric" term. Since $L=1$, the nucleon component $[{\bf 70},
^2 8, 2, 0, 1/2 ]$ can be excited to $J=1/2$ and 3/2 corresponding
to $P_{11}$ and $P_{13}$ in $[{\bf 20}, ^2 8, 2, 1, J]$. The
helicity amplitudes (not including the mixing angle) are presented
in Table~\ref{tab-1} with the spatial integral
\bea
A  &=& 6\sqrt{\frac{\pi}{k_0}}\mu_0\frac{1}{g} \langle\psi_{211}^a
| e^{ik r_{3z}}p_{3+}|\psi_{200}^\rho\rangle\nonumber\\
&=&-6\sqrt{\frac{\pi}{k_0}}\mu_0\frac{1}{g}
\times\frac{2k}{3\sqrt{3}}e^{-k^2/6\alpha_h^2}.
\eea
These amplitudes may be compared with those for $\gamma p\to
D_{15}$ as listed in Table~\ref{tab-1}, for which the spatial
integral is
\bea
B&=&6\sqrt{\frac{\pi}{k_0}}\mu_0 k\langle \psi_{110}^\rho |
e^{ik r_{3z}}| \psi_{200}^\rho\rangle\nonumber\\
&=&6\sqrt{\frac{\pi}{k_0}}\mu_0 k\times
(-i)\frac{k}{3\alpha_h}e^{-k^2/6\alpha_h^2}.
\eea
Additional $P_{11}$ and $P_{13}$ from representation {\bf 20}
automatically raise questions about the quark model assignments of
the observed $P_{11}$ and $P_{13}$ states, among which
$P_{11}(1440)$, $P_{11}(1710)$, and $P_{13}(1720)$ are
well-established resonances, while signals for $P_{13}(1900)$ and
$P_{11}(2100)$ are quite poor~\cite{pdg2004}.


\section{Positive parity $N^*$ up to 2 GeV }

At $N=2$ in the quark model a quark-diquark spectrum allows $[{\bf
56}, 0^+]$, $[{\bf 56}, 2^+]$, and $[{\bf 70}, 0^+]$$[{\bf 70},
2^+]$. If all $qqq$ degrees of freedom can be excited,
correlations corresponding to $[{\bf 20}, 1^+]$ are also possible.

This implies the following $N^*$ and $\Delta$ states (the
superscripts denoting the $qqq$ net spin state as $2S+1$):
\begin{eqnarray}
{[\bf 56},0^+] & =& P_{11}(^2N); P_{33}(^4\Delta) \nonumber\\
{[\bf 56},2^+] & =& P_{13},F_{15}(^2N);
P_{31},P_{33},F_{35},F_{37}(^4\Delta) \nonumber\\
{[\bf 70},0^+]
& =& P_{11}(^2N); P_{13}(^4N); P_{31}(^2\Delta) \nonumber\\
{[\bf 70},2^+] &= & P_{11},P_{13},F_{15},F_{17}(^4N);
P_{13},F_{15}(^2N);P_{33},F_{35}(^2\Delta) .
\end{eqnarray}
Without the {\bf 20}-plets, a commonly accepted scheme is as
follows \cite{capstick-roberts,ik70mix}.

The $P_{11}(1440)$ is assigned to $[{\bf 56}, 0^+]$ at $N=2$ with
$P_{33}(1660)$ as its isospin 3/2
partner~\cite{roper,dalitz-moorhouse}. The photoproduction
amplitudes off proton and neutron, $A^p_{1/2} = -0.065 \pm 0.004$
and $ A^n_{1/2} = +0.040 \pm 0.010$ (GeV)$^{-1/2}$, are consistent
with M1 transitions in ratio $-3:2$ as for the
nucleons~\cite{pdg2004,cko,fkr}. The mass splitting of the
$N(1440)-\Delta(1660)$ is consistent with the hyperfine splitting.

For the [{\bf 56},$2^+$] the $F_{37}(1950)$ is uniquely assigned.
The hyperfine mass splitting naturally associates the $F_{15}(1680)$ as
a partner; this is further confirmed by its photoproduction amplitudes,
which satisfy selection rules from both proton and neutron
targets~\cite{cko,fkr}. The $P_{13}(1720)$ is prima facie
associated with the $F_{15}(1680)$ but the photoproduction
amplitudes do not easily fit: the helicity 3/2 amplitude of the
$P_{13}$ is predicted to be one-half that of the $F_{15}$, which does not fit well
with the data~\cite{pdg2004}.
The overall message from the {\bf 56} is that the constituent quark degrees of freedom are
manifested at $N=2$ even though $N\pi$ couplings are strong as
evidenced by $\Gamma_T \sim 350$ MeV for $P_{11}(1440)$. This sets
the challenge of assessing the situation for the rest of the $N=2$
levels.

Only three $P_{11}$ states are expected at $N=2$ if the {\bf
20}-plets cannot be excited. As to assigning these states: (i) The $P_{11}(1440)$ has already been
assigned to {\bf 56}; (ii) the $P_{11}$ of $[{\bf 70}, ^4 8; 2^+]$
is expected to be heavier\cite{capstick-roberts,ik70mix} and the
signals for $F_{17}(1990)$, $F_{15}(2000)$, $P_{13}(1900)$ and
$P_{11}(2100)$~\cite{pdg2004} lead to a plausible assignment of
these four in $[{\bf 70}, ^4 8, 2, 2, J]$; (iii) this leaves
$P_{11}(1710)$ a natural candidate for $[{\bf 70}, ^2 8, 2, 0,
1/2]$. Nonetheless, there are still states of $[{\bf 70}, ^2 8,
2,2,J]$ missing, and another $P_{13}$ and $F_{15}$ are needed.

Within the approximations that excitation and decays are dominated
by single quark transition, and that there is no mixing among the
{\bf 56} and {\bf 70} basis states, the assignments of $P_{13}$
and $F_{15}$ can be determined. The Moorhouse and $\Lambda$
selection rules give the following filters for these states:
$P_{13}(^4N)$ in [{\bf 70},$0^+$] decouples from $K\Lambda$ and
$\gamma p$ but couples to $\gamma n$. This contrasts with
$P_{13}(^2N)$ in either of [{\bf 70},$2^+$] or [{\bf 56},$2^+$]
which couple to all three channels. The helicity 3/2 amplitude of
 $F_{15}$ in either of the [{\bf 70},$2^+$] or [{\bf 56},$2^+$]
couples to photons with amplitude that is twice the size of its
$P_{13}$ counterparts.

Table~\ref{tab-2} shows the photoproduction amplitudes expected.
The $P_{11}(1710)$ fits with $[{\bf 70}, ^2 8, 2, 0, 1/2]$ due to
the large error bars, however neither of the $P_{13}(1710/1900)$
fit easily with being pure {\bf 56} or {\bf 70} states. When the
QCD mixing effects are included the agreement improves, in that
small couplings of $^4N$ states to $\gamma p$ are predicted, in
accord with data. However, the implication is the added complexity
that an additional $P_{11}$ and $P_{13}$ correlation in [{\bf
20},$1^+$] is allowed. Most immediately this prevents associating
the $P_{11}(1710)$ as [{\bf 70,$0^+$}] simply on the grounds of
elimination of alternative possibilities. Thus we now consider
what are the theoretical signals and what does experiment
currently say.

Qualitatively one anticipates $P_{13}(^4N)$ having a small but
non-zero coupling to $\gamma p$, the $\gamma n$ being larger while
the $K \Lambda$ decay is still forbidden. For the {\bf 20} states
$P_{11,13}(^2N)$ both $\gamma p$ and $\gamma n$ amplitudes will be
small and of similar magnitude. However, mixing with their
counterparts in {\bf 56} and {\bf 70} may be expected. In
Table~\ref{tab-2}, we list the helicity amplitudes for the
$P_{11}(1710)$, $P_{13}(1720)$ and $P_{13}(1900)$ with all the
possible quark model assignments and the mixing angles from
Eq.~(\ref{nmix}). The amplitudes for the $P_{11}$ and $P_{13}$ of
$[{\bf 20}, ^2 8, 2,1, J]$ are the same order of magnitude as the
Moorhouse-violating $\gamma p\to D_{15}(1675)$ in
Table~\ref{tab-1}.  For the $P_{11}(1710)$ all three possible
configurations  have amplitudes compatible with experimental data.
For $P_{13}(1720)$, assignment in either $[{\bf 56}, ^2 8; 2^+]$
or $[{\bf 70}, ^2 8; 2^+]$ significantly overestimates the
data~\cite{close-li,li-close} for $A^p_{1/2}$ if it is a pure
state.  Table~\ref{tab-2} shows that the presence of {\bf 20}
cannot be ignored, should be included in searches for so-called
``missing resonances", and that a possible mixture of the {\bf
20}-plets  may lead to significant corrections to the results
based on the conventional {\bf 56} and {\bf 70}.

As a benchmark for advancing understanding we propose the
following scenario, as a challenge for experiment:
can one eliminate the {\it extreme} possibility that
$P_{11}(1710)$ and $P_{13}(1720)$ are consistent with being in
${\bf 20}$ configurations? There are already qualitative indications that they
are not simply {\bf 56} or {\bf 70}. Their hadronic decays differ noticeably
from their sibling $P_{11}(1440)$: compared with the
$P_{11}(1440)$ in {\bf 56} for which $\Gamma_T \sim 350 $ MeV with
a strong coupling to $N\pi$, their total widths are $\sim 100$
MeV, with $N\pi$ forming only a small part of this.

Consider now the phenomenology were these states in {\bf 20}. As
${\bf 20} \nrightarrow {\bf 56} \otimes {\bf 35}$, whereas ${\bf
20} \rightarrow {\bf 70} \otimes {\bf 35}$ is allowed, decay to $N
\pi$ will be allowed only through the ${\bf 70}$ admixtures in the
nucleon. Using the wavefunction in Eq.~(\ref{nmix}), this implies
that $N\pi$ widths will be suppressed by an order of magnitude
relative to allowed widths such as from {\bf 56} or {\bf 70}
initial states. Decays ${\bf 20}\to N^*({\bf 70})+\pi \to N
\pi\pi$ or $N \eta \pi$ can occur in leading order when phase
space allows. These results make the $P_{11}(1710)$ and
$P_{13}(1720)$ extremely interesting as not only are their total
widths significantly less than the $P_{11}(1440)$ but the dominant
modes for $P_{11}(1710)$ and $P_{13}(1720)$ are to $N\pi\pi$,
which allows a possible cascade decay of  ${\bf 20}\to N^*({\bf
70})\pi\to N\pi\pi$. In particular, $B.R._{P_{13}(1720)\to
N\pi\pi}\sim 70\%$ could contain significant $N^*({\bf 70})\pi$;
present partial wave analysis has not included this
possibility~\cite{manley}. For $P_{11}({\bf 20})$ the $S$-wave
decay $P_{11} \to S_{11} \pi$ is expected to be a dominant
two-body decay; hence we urge a search for $P_{11} \to N \eta
\pi$. Decays to ${\bf 56} + M$ (where $M$ denotes a meson) will
occur through mixing, as for photoproduction, and so $K\Lambda$
may be expected at a few percent in branching ratio. In general we advocate
that partical wave analyses of $N\pi\pi$ allow for the possible
presence of $N^*$({\bf 70})$\pi$.

In order to classify these states, we advocate that partial wave
analysis quantifies the couplings to $N^*(\bf 70)\pi$ and also
includes data on $K\Lambda$ and/or $K^*\Lambda$ relative to $K
\Sigma$ as the $\Lambda$ selection rule can then be brought to
bear. The $\Lambda$ selection rule is useful for classifying the
$P_{11}$ and $P_{13}$ in either $[{\bf 56}, ^2 8]$ and $[{\bf 70},
^2 8]$, or $[{\bf 70}, ^4 8]$ and $[{\bf 20}, ^2 8]$, by looking
at their decays into $K \Lambda$ and/or $K^*\Lambda$. The
Moorhouse selection rule can distinguish $[{\bf 70}, ^4 8]$ and
$[{\bf 20}, ^2 8]$ since the $[{\bf 70}, ^4 8]$ will be suppressed
in $\gamma p$ but sizeable in $\gamma n$, while the $[{\bf 20}, ^2
8]$ will be suppressed in both. The $[{\bf 70}, ^4 8]$ decays to
$K\Lambda$ will be suppressed relative to $K \Sigma$ for both
charged and neutral $N^*$. $J/\psi \to \bar{p} + N^*$ is a further
probe of $N^*$ assignments, which accesses {\bf 56} in leading
order and {\bf 70} via mixing while {\bf 20} is forbidden. Hence
for example $J/\psi \to \bar{p} + (P_{11}:P_{13})$ probes the {\bf
56} and {\bf 70} content of these states. Combined with our
selection rule this identifies $J/\psi \to \bar{p} + K \Lambda$ as
a channel that selects the {\bf 56} content of the $P_{11}$ and
$P_{13}$.

In summary, with interest in $N^*$ with masses above 2 GeV coming
into focus at Jefferson Laboratory and accessible at BEPC with
high statistic $J/\psi\to \bar{p}+ N^*$, this new selection rule
should be useful for classifying baryon resonances and
interpreting $\gamma N \to K\Lambda$, $K^*\Lambda$, $K\Sigma$ and
$K^*\Sigma$~\cite{hleiwaqi,guo}. A coherent study of these
channels may provide evidence on the dynamics of diquark
correlations and the presence of {\bf 20}-plets, which have
hitherto been largely ignored.

\section*{Acknowledgement}

We are grateful to G. Karl for helpful comments. This work is
supported, in part, by grants from the U.K. Engineering and
Physical Sciences Research Council (Grant No. GR/S99433/01), and
the Particle Physics and Astronomy Research Council, and the
EU-TMR program ``Eurodice'', HPRN-CT-2002-00311, and the Institute
of High Energy Physics, Chinese Academy of Sciences.

%
\begin{table}[ht]
\begin{tabular}{|c|cc|}
\hline Final state & \ \ \ \ \ $A^p_{1/2} $  \ \ \ \ \ & \ \ \ \ \
$A^p_{3/2}$ \ \ \
\\[1ex]
\hline $P_{11}$ & $-\frac{1}{3\sqrt{3}} A $ & -
\\[1ex]
$P_{13}$ & $\frac{1}{3\sqrt{6}} A$ & $\frac{1}{3\sqrt{2}} A $
\\[1ex]\hline
$D_{15}$ & $\frac{1}{3}\sqrt{\frac{1}{15}} B$ &
$\frac{1}{3}\sqrt{\frac{2}{15}} B$ \\[1ex]
Theo.  & 14 & 20 \\[1ex]
Exp.  & $19\pm 8$ & $15\pm 9$ \\[1ex]\hline
\end{tabular}
\caption{Helicity amplitudes for $[70, ^2 8, 2,0, 1/2]\to [20, ^2
8, 2, 1, J]$ for $P_{11}$ and $P_{13}$, and $[70, ^2 8, 2,0,
1/2]\to [70, ^4 8, 1, 1, J]$ for $D_{15}$. $A$ and $B$ are the
corresponding spatial integrals. The theory calculations with the
mixing angle and experimental data~\cite{pdg2004} for $\gamma p\to
D_{15}(1675)$ are also listed with unit $10^3\times$GeV$^{1/2}$. }
\label{tab-1}
\end{table}


\begin{table}[ht]
\begin{tabular}{|c|c|ccccccc|}
\hline Reso. & Heli. amp. & $[{\bf 56}, ^2 8; 2^+]$ & $[{\bf 70},
^2 8; 0^+]$ & $[{\bf 70}, ^28; 2^+]$ & $[{\bf 70}, ^48; 0^+]$ &
$[{\bf
70}, ^48; 2^+]$ & $[{\bf 20}, ^28; 1^+]$ & Exp. data \\[1ex]\hline
$P_{11}(1710)$ & $A^p_{1/2}$ & * & 32 & * & * & $-8$ & $-15$ &
$+9\pm 22$ \\[1ex]\hline
$P_{13}(1720)$ & $A^p_{1/2}$ & 100 & * & $-71$ & 17 & 7 & $-11$ &
$+18\pm 30$ \\[1ex]
& $A^p_{3/2}$ & 30 & * & $-21$ & 29 & 12 & $-18$ & $-19\pm 20
$\\[1ex]\hline
$P_{13}(1900)$ & $A^p_{1/2}$ & 110 & * & $-78$ & 14 & 9 & $-10$ &
$-17^*$ \\[1ex]
& $A^p_{3/2}$ & 39 & * & $-28$ & 24 & 16 & $-17$ & $+31^*$\\[1ex]\hline
\end{tabular}
\caption{Helicity amplitudes for the $P_{11}(1710)$ and
$P_{13}(1720)$ with all the possible quark model assignments for
them. The data are from PDG~\cite{pdg2004}, and numbers have a
unit of $10^3\times$GeV$^{1/2}$. The numbers with ``*" are from
multichannel studies~\cite{penner-mosel}. } \label{tab-2}
\end{table}

\end{document}